\documentclass[10pt,letterpaper,english,reprint,superscriptaddress]{revtex4-1}
\usepackage{mathptmx}
\usepackage[T1]{fontenc}
\usepackage[latin9]{inputenc}
\usepackage{color}
\usepackage{textcomp}
\usepackage{amsmath}
\usepackage{amssymb}
\usepackage{graphicx}

\makeatletter

\pdfpageheight\paperheight
\pdfpagewidth\paperwidth

\usepackage{amssymb}
\renewcommand{\fnum@figure}{\textbf{Figure~\thefigure}}

\setcitestyle{super}

\makeatother

\usepackage{babel}
\begin{document}
\title{\textcolor{black}{Observation of an antiferromagnetic quantum critical
point in high-purity LaNiO$_{3}$}}
\author{Changjiang Liu}
\affiliation{{\small{}Materials Science Division, Argonne National Laboratory,
Lemont, IL 60439}}
\author{Vincent F. C. Humbert}
\affiliation{{\small{}Department of Physics, University of Illinois at Urbana-Champaign,
Urbana, IL 61801, USA}}
\author{Terence Bretz-Sullivan}
\affiliation{{\small{}Materials Science Division, Argonne National Laboratory,
Lemont, IL 60439}}
\author{Gensheng Wang}
\affiliation{{\small{}High Energy Physics Division, Argonne National Laboratory,
Lemont, IL 60439, USA}}
\author{Deshun Hong}
\affiliation{{\small{}Materials Science Division, Argonne National Laboratory,
Lemont, IL 60439}}
\author{Friederike Wrobel}
\affiliation{{\small{}Materials Science Division, Argonne National Laboratory,
Lemont, IL 60439}}
\author{Jianjie Zhang}
\affiliation{{\small{}High Energy Physics Division, Argonne National Laboratory,
Lemont, IL 60439, USA}}
\author{Jason D. Hoffman}
\affiliation{{\small{}Department of Physics, Harvard University, Cambridge, MA
02138, USA}}
\author{John E. Pearson}
\affiliation{{\small{}Materials Science Division, Argonne National Laboratory,
Lemont, IL 60439}}
\author{J Samuel Jiang}
\affiliation{{\small{}Materials Science Division, Argonne National Laboratory,
Lemont, IL 60439}}
\author{Clarence Chang}
\affiliation{{\small{}High Energy Physics Division, Argonne National Laboratory,
Lemont, IL 60439, USA}}
\author{Alexey Suslov}
\affiliation{{\small{}National High Magnetic Field Laboratory, Tallahassee, FL
32310, USA}}
\author{Nadya Mason}
\affiliation{{\small{}Department of Physics, University of Illinois at Urbana-Champaign,
Urbana, IL 61801, USA}}
\author{M. R. Norman}
\affiliation{{\small{}Materials Science Division, Argonne National Laboratory,
Lemont, IL 60439}}
\author{Anand Bhattacharya}
\email{anand@anl.gov}

\affiliation{{\small{}Materials Science Division, Argonne National Laboratory,
Lemont, IL 60439}}
\begin{abstract}
\textcolor{black}{Amongst the rare-earth perovskite nickelates, LaNiO$_{3}$
(LNO) is an exception. While the former have insulating and antiferromagnetic
ground states, LNO remains metallic and non-magnetic down to the lowest
temperatures. It is believed that LNO is a strange metal, on the verge
of an antiferromagnetic instability. Our work suggests that LNO is
a quantum critical metal, close to an antiferromagnetic quantum critical
point (QCP). The QCP behavior in LNO is manifested in epitaxial thin
films with unprecedented high purities. We find that the temperature
and magnetic field dependences of the resistivity of LNO at low temperatures
are consistent with scatterings of charge carriers from weak disorder
and quantum fluctuations of an antiferromagnetic nature. Furthermore,
we find that the introduction of a small concentration of magnetic
impurities qualitatively changes the magnetotransport properties of
LNO, resembling that found in some heavy-fermion Kondo lattice systems
in the vicinity of an antiferromagnetic QCP.}
\end{abstract}
\maketitle
\noindent \textbf{Introduction}

\noindent In the vicinity of a quantum phase transition (QPT) near
$T=$ 0 K, a material can be tuned in and out of an ordered state
using a parameter other than temperature, such as magnetic field or
pressure \citep{Coleman2005,Gegenwart2008}. Near a QPT, quantum fluctuations
of the order parameter can have profound influence on the properties
of the material, out to high temperatures \citep{Sachdev2009}. For
example, in a metal, quantum fluctuations can introduce long-range
interactions between mobile electrons causing a breakdown of the Landau
F\textcolor{black}{ermi liquid (LFL),\citep{Schofield1999} leading
to a strange metal with anomalous transport and thermodynamic properties
\citep{Stewart2001}. Quantum fluctuations can also mediate superconducting
pairing of carriers and give rise to unusual responses to electric
and magnetic fields. In rare instances, a material can be found to
be intrinsically quantum critical, perched on the edge of a QPT without
the need for tuning. Here we report on signatures of an antiferromagnetic
QCP in high-purity LaNiO$_{3}$ thin films. We find that the resistivity
$\rho(T)$ shows a linear temperature dependence over almost a decade
of $T$ below \textasciitilde{} 1.1 K in our cleanest samples. The
linear-in-$T$ resisti}vity crosses over to a $T^{2}$ dependence
in a magnetic field, consistent with the presence of antiferromagnetic
quantum critical fluctuations.

The rare-earth nickelates $(Re$-NiO$_{3})$ are a widely studied
family of materials that display rich electronic and magnetic properties
as a result of competition between itinerancy and electron-electron
and electron-lattice interactions that tend to localize carriers and
give rise to an antiferromagnetic (AFM) insulating ground state. They
have a perovskite structure, with Ni cations at the center of corner-sharing
O octahedra. For smaller \emph{Re} cations \citep{Torrance1992,Catalano2018},
an insulating state is obtained at temperatures below $T_{\mathrm{MIT}}$
(metal-insulator transition temperature), accompanied by a structural
distortion. At yet lower temperatures $T_{\mathrm{N}}\leq T_{\mathrm{MIT}}$
$(T_{\mathrm{N}}$ is the N$\mathrm{\acute{e}}$el temperature), an
AFM state is obtained. As the \emph{Re} cation radius increases, $T_{\mathrm{MIT}}$
decreases, merges with $T_{\mathrm{N}}$, and eventually both are
driven to zero. LaNiO$_{3}$ (LNO), with the largest \emph{Re} cation,
is the only $Re$-NiO$_{3}$ nickelate that is metallic down to the
lowest temperatures. LNO is a correlated metal \citep{Zhou2014} and
it has long been suspected that the properties of LNO are influenced
by its proximity to magnetic and structural instabilities \citep{Hoffman2016,Fabbris2018},
perhaps even by quantum fluctuations resulting from these instabilities
\citep{Allen2015,Subedi2018}. However, clear experimental evidence
for the effect of quantum fluctuations at low temperatures has not
been reported in LNO until now. 

Here, we report on systematic study on a series of LNO samples with
varying degrees of disorder. The dependence of the resistivity on
temperature show agreement with theories that consider the interplay
between scattering from disorder and quantum AFM fluctuations. Furthermore,
we observe signatures of spin-flip scattering from localized spins
in samples with greater levels of disorder at higher temperatures,
while at low temperatures signatures of antiferromagnetic correlations
emerge, similar to behavior observed in some heavy fermion systems.
Our findings indicate that the low temperature transport properties
of LNO arise from the interplay between an antiferromagnetic quantum
critical point, impurity scattering and the interaction of itinerant
carriers with localized spins.

\noindent \textbf{Results}

\noindent \textbf{Sample preparation}. Epitaxial LNO thin films are
grown on (001) oriented $\mathrm{(LaAlO_{3})_{0.3}(Sr_{2}AlTaO_{6})_{0.7}}$
(LSAT) substrates by ozone-assisted molecular beam expitaxy. The growth
parameters for controlling the La/Ni ratio are determined from Rutherford
backscattering spectrometry (RBS) measurement (see Supplementary Figure
1). To ensure that the oxygen vacancies are minimized, we used a high
ozone flux with a background pressure of $7\times10^{-6}$ torr. The
growth process was monitored by by reflection high-energy electron
diffraction (RHEED) (see Supplementary Figure 2). Details of the sample
growth, the control of stoichiometry and X-ray characterizations are
presented in the Methods section and Supplementary Figure 3. Transport
measurements were performed on six-terminal devices patterned in Hall
bar geometry using photolithography. Seven samples are studie\textcolor{black}{d
in this work, and they are labeled as LNO\_\#, with \# being close
to the residual resistivity ratio {[}RRR = $\rho$(300 K)/$\rho$(2
K){]} of the sample. The disorder level in the sample may be characterized
approximately by the residual resistivity or RRR of the sample (note
that Matthiessen's rule might not apply at $T$ = 2 K due to an interplay
between disorder and other scattering mechanisms).}

\begin{figure*}[!tp]
\includegraphics[width=16cm]{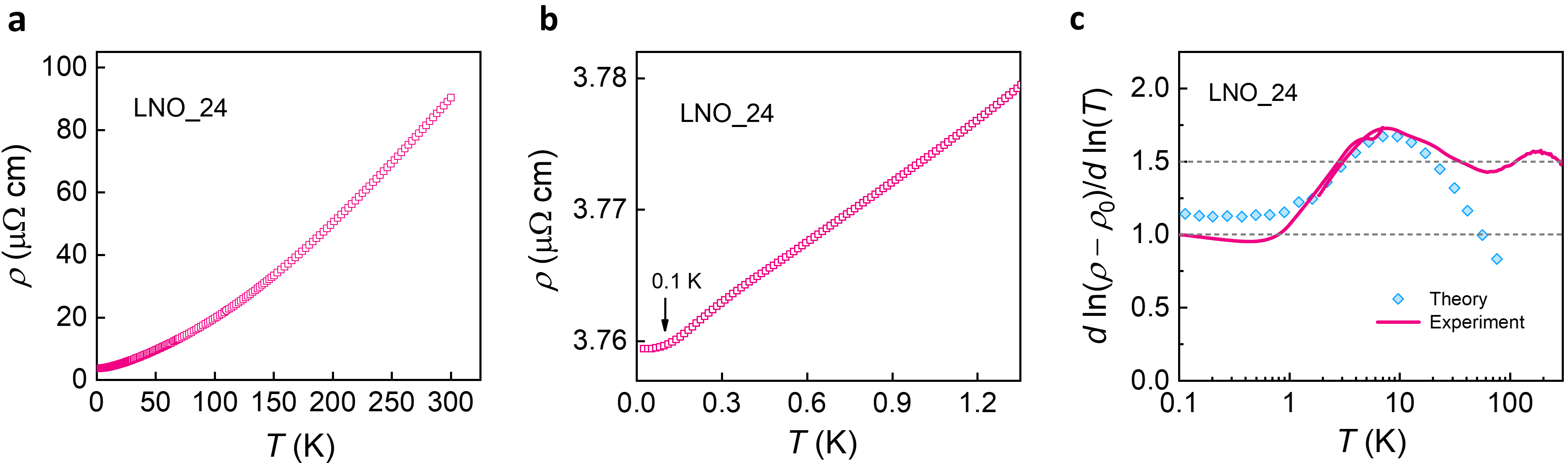}\caption{\textbf{Temperatu}\textbf{\textcolor{black}{re dependence of resistivity
and evolution of the resistivity exponent.}}\textcolor{black}{{} }\textbf{\textcolor{black}{a}}\textcolor{black}{{}
Resistivity of LNO measured as a function of temperature from 300
K to 2 K. }\textbf{\textcolor{black}{b}}\textcolor{black}{{} Linear-in-temperature
resistivity observed at temperatures below about 1.1 K. }\textbf{\textcolor{black}{c}}\textcolor{black}{{}
Temperature dependence of the resistivity exponent $\alpha$ (solid
line) from 300 K to 100 mK. The two horizontal dashed lines indicate
exponent values of 1 and 1.5, respectively. Symbols show theoretical
predictions of the resistivity exponent $\alpha$ for a clean sample
($k_{F}l$\textasciitilde{} 1000) at AFM QCP.}}
\end{figure*}

\noindent \textbf{Resistivity measurement in the high-purity sample}\textcolor{black}{.
Figure 1a shows the temperature dependence of the resistivity measured
on sample LNO\_24. This sample shows a resistivity of 3.8 $\mathrm{\mathrm{\mu\varOmega}}$
cm and a mobility of about 160 $\mathrm{cm^{2}}$ V$^{-1}$ s$^{-1}$
at $T=2$ K. The corresponding RRR is about 24, which is so far the
highest RRR reported for LNO (see Supplementary Note 1 and Table 1).
When the transport measurement was extended down to 25 mK, the resistivity
continues to decrease without showing any flattening until about 100
mK, and in fact it shows an unexpected linear-in-temperature dependence
in the temperature range 0.1 K < $T$ < 1.1 K (Fig. 1b). The same
behavior in the resistivity is observed in multiple samples (see Supplementary
Figure 4). A description of the uncertainties of the data points is
presented in the Methods section. Phonons are nominally not relevant
in this low temperature regime given that the Debye temperature of
LNO is above 400 K.\citep{Zhang2017} Ni-O bond length fluctuations
could be a source of linear$-T$ resistivity at low temperatures \citep{Rivadulla2003},
and there may be other mechanisms as well (charge fluctuations, Umklapp
scatterings). However, these mechanisms are less natural for explaining
our data as they would have a weak dependence on magnetic field, which
will be discussed in the following. This behavior of the resistivity
in LNO is in sharp contrast with the LFL theory, which predicts a
quadratic temperature dependence of resistivity at low temperatures.
In Fig. 1c, we plot the resistivity exponent $\alpha$ (solid line)
in $\rho(T)=\rho_{0}+AT^{\alpha}$, as a function of temperature.
Here $\rho_{0}$ is t}he residual resistivity, and $\alpha$ is calculated
from $d\ln(\rho-\rho_{0})/d\ln(T)$. Before taking derivatives, the
raw data was first smoothed using a B-Spine method. We find that $\alpha$
initially oscillates around a value of 1.5 from $T=300$ K to about
30 K (Fig. 1c). As temperature decreases further, $\alpha$ first
increases to a value of about 1.75 near $T$ = 7.5 K and then begins
to decrease and approaches a constant value of \textasciitilde{} 1
in the sub-Kelvin regime.

Previous studies found that the resistivity of LNO shows a $T^{1.5}$
power law at temperatures above about 30 K,\citep{Zhou2014,Zhang2017,Guo2018}
and anomalous exponents have been observed in other nickelates \citep{Liu2013,Mikheev2015}.
Similar behavior is observed here as we saw in Fig. 1c (see also Supplementary
Figure 5). A $T^{1.5}$ power law scaling is often attributed to scattering
by AFM spin fluctuations, as proposed by the Hertz--Millis--Moriya
model for systems close to an AFM QCP.\citep{moriya2012spin,Millis1993,LohneyesenRMP2007}
Electronic states on the Fermi surface connected by wave vectors of
the incipient AFM ordering (hot regions in $k$-space) are strongly
scattered. However, these hot regions only occupy a finite phase space
on the Fermi surface, and in the clean limit they are shorted out
by the cold regions where such scattering does not operate \citep{Hlubina1995}.
It was realized subsequently that the scattering due to quantum AFM
spin fluctuations can lead to a peculiar evolution \citep{Rosch1999,Rosch2000}
of $\alpha$ with \emph{T} that depends on the level of disorder.
For systems close to AFM QCP in the dirty limit, a $T^{1.5}$ power
law should be observed over a large range of temperature. For cleaner
systems ($k_{\mathrm{F}}l>10$, where $k_{\mathrm{F}}$ is the Fermi
wavevector and $l$ is the mean free path), as temperature decreases,
$\alpha$ first increases and approaches values close to 2 (symbols
in Fig. 1c), as expected for LFL quasiparticles. At lower temperatures,
$\alpha$ decreases continuously to a value \textasciitilde{} 1. A
signature characteristic of this model is a bump in $\alpha$ as a
function of \emph{T,} \citep{Rosch1999} which is clearly observed
here at temperature $T_{\mathrm{Bump}}\sim$ 7.5 K (Fig. 1c). At even
lower temperatures T < $10^{-3}$ $T_{\mathrm{Bump}}$, which are
not experimentally accessible for us, $\alpha$ increases again to
the dirty limit value of 1.5 according to this theory. \textcolor{black}{Using
the resistivity and Hall measurements (see Supplementary Figure 6),
we estimate $k_{\mathrm{F}}l$ \textasciitilde{} 500 for our cleanest
samples at $T$ = 2 K.} Symbols in Fig. 1c are calculations reproduced
from ref. 22 for a clean sample with $k_{\mathrm{F}}l\sim1000$ (see
also Supplementary Figure 7). \textcolor{black}{Further details of
the analysis are presented in Supplementary Note 2.}

\begin{figure}[b]
\includegraphics[width=8cm]{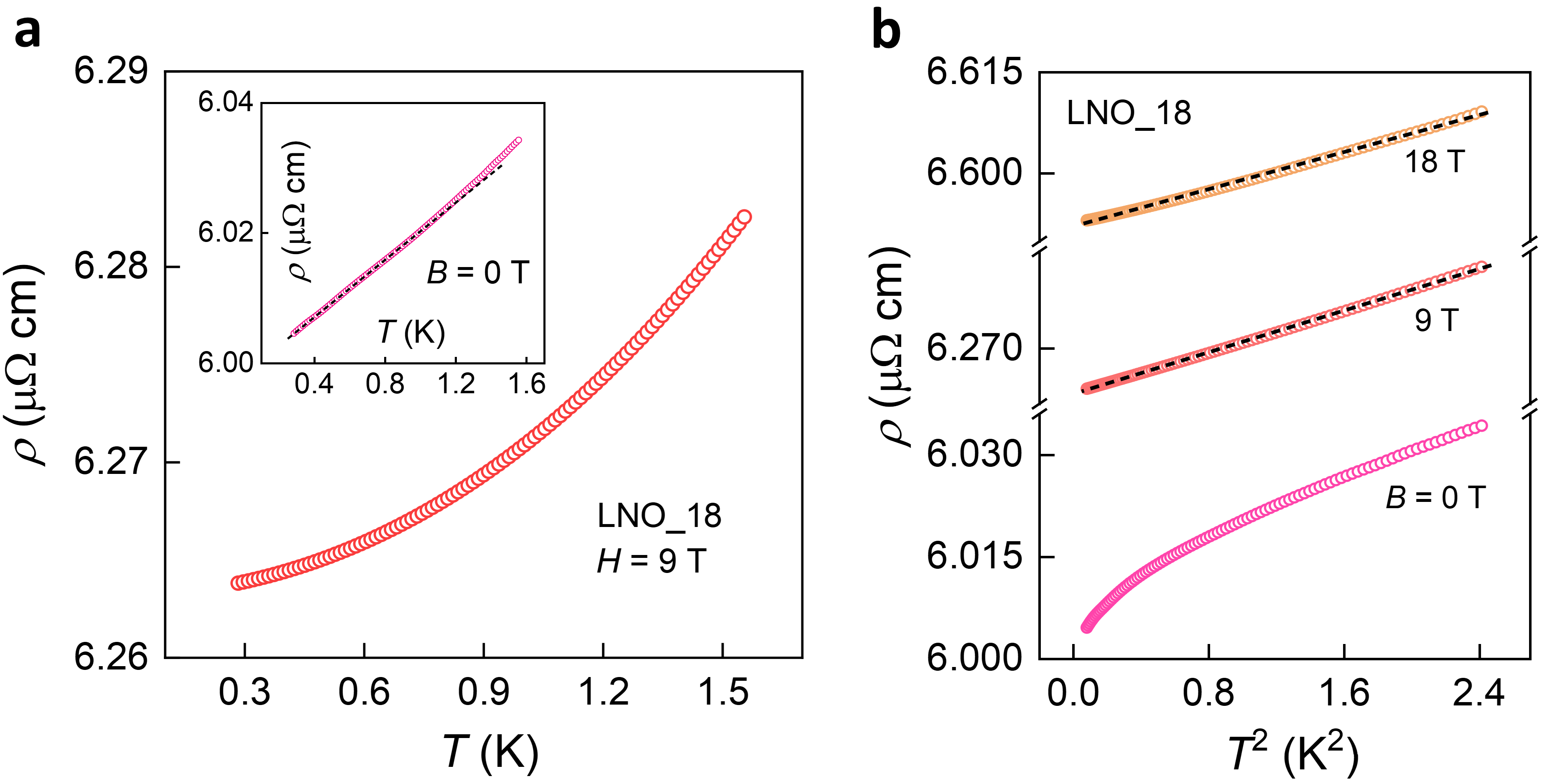}\caption{\textbf{Temperatu}\textbf{\textcolor{black}{re dependence of resistivity
under a magnetic field.}}\textcolor{black}{{} }\textbf{\textcolor{black}{a}}\textcolor{black}{{}
Resistivity measurement under a magnetic field of 9 tesla for LNO\_18.
Inset shows the linear-in-temperature behavior under zero field. }\textbf{\textcolor{black}{b}}\textcolor{black}{{}
Temperature dependence of resistivity plotted versus the square of
temperature for different magnetic fields. The measurements at $B=9$
and 18 tesla show a linear dependence on $T^{2}.$ Note the positive
magnetoresistance. The standard deviation of the mean at each point
is smaller than the symbol size.The black dashed lines in (}\textbf{\textcolor{black}{b}}\textcolor{black}{)
and the inset of (}\textbf{\textcolor{black}{a}}\textcolor{black}{)
are straight guidelines. }}
\end{figure}

\begin{figure*}[!tp]
\includegraphics[width=16cm]{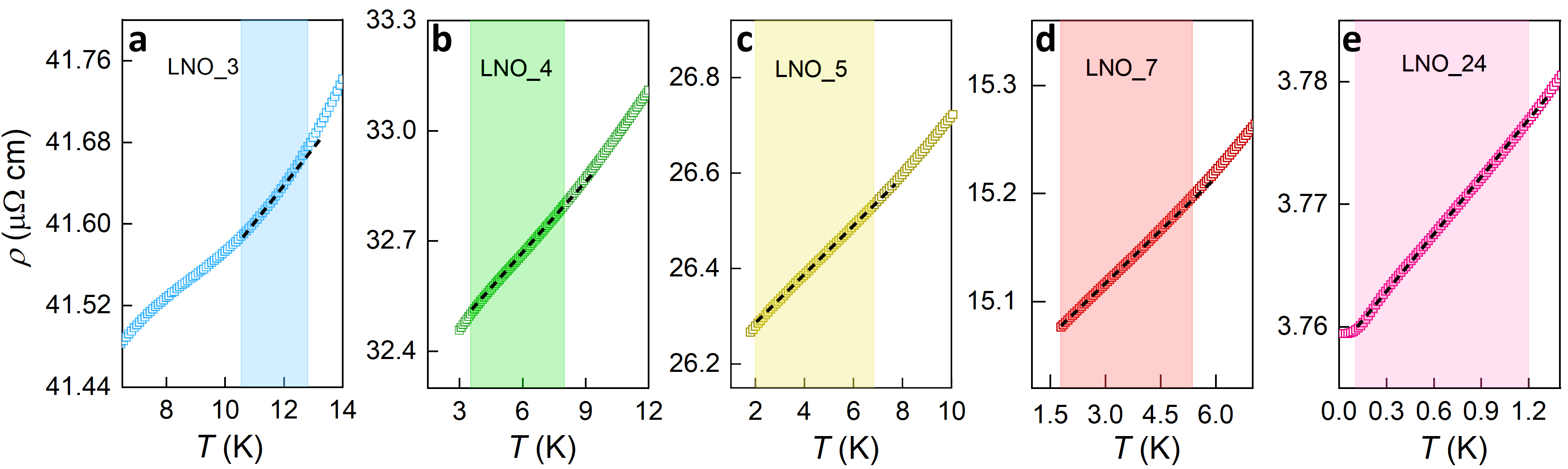}\textcolor{black}{\caption{\textbf{Resistivity of samples with different disorder levels.} \textbf{\textcolor{black}{a}}\textcolor{black}{-}\textbf{\textcolor{black}{e}}\textcolor{black}{{}
}Low-temperature measurement of resistivity on samples with different
impurity levels. The shaded region in the plot indic\textcolor{black}{ates
the temperature range where the resistivity show a linear temperature
dependence (quasilinear for low RRR samples). The black dashed lines
are straight guidelines. A downward curvature in the resistivity is
observed in these samples near the low temperature end $T\protect\leq T_{\mathrm{low}}$. }}
}
\end{figure*}

\noindent \textbf{Restoration of LFL under a magnetic field}. If AFM
fluctuations ne\textcolor{black}{ar a QCP are responsible for the
linear $\rho(T)$ that we observe at low temperatures, an applied
magnetic field may be used to suppress these fluctuations and restore
the LFL. Such behavior has been observed in heavy fermion compounds
close to a QCP - for example in $\mathrm{YbRh_{2}(Si_{1-x}Ge_{x})_{2}}$
and $\mathrm{CeCoIn_{5}}$. \citep{Custers2003,Paglione2003} Figure
2a shows our measurement results on LNO\_18 under a magnetic field
of 9 T, presenting a quadratic dependence of resistivity on temperature,
i.e., $\rho(T)=\rho_{0}+AT^{2}$. Inset of Fig. 2a shows the linear
$\rho(T)$ in zero magnetic field in the sub-Kelvin regime, as has
b}een observed in other clean samples. Thus, a restoration of LFL
behavior in LNO is observed under a magnetic field. This is seen more
clearly in Fig. 2b\textcolor{black}{, where we plot $\rho(T)$ vs
$T^{2}$ at different fields. The data can be fit well to a $T^{2}$
dependence for $B$ = 9 T and 18 T, with corresponding $A$ coefficients
in $\rho(T)$ being $8\times10^{-3}$ and $7\times10^{-3}$ $\mathrm{\mu}\Omega$
cm $\mathrm{K}^{-2}$, respectively. These values of $A$ are about
3 to 4 times larger than those, $\sim2\times10^{-3}$ $\mathrm{\mu}\Omega$
cm $\mathrm{K}^{-2}$, reported for LNO previously \citep{Son2010,Zhang2017,Guo2018},
which is suggestive of enhanced scattering near the QCP. We note that
this crossover to LFL in the resistivity under a magnetic field is
seen only in samples with RRR $\geq$ 18. Therefore, the presence
of disorder in LNO could detune the system away from the QCP.}

\noindent \textbf{Interplay of AFM spin fluctuations}\textcolor{black}{{}
}\textbf{and disorder}\textcolor{black}{. To examine the role of disorder
on the quantum critical behavior of LNO, we grew a series of samples
under different ozone pressure, substrate temperatures, and with slight
variance in La/Ni ratio (< 2\%). The growth conditions of each sample
are presented in Supplementary Table 2. The main form of impurities
introduced during the growth are oxygen vacancies and $\mathrm{Ni^{2+}},$
which also act as local magnetic moments. The sample with the highest
RRR was grown under the highest effective ozone pressure, a substrate
temperature of 615 \textcelsius , and a La/Ni ratio very close to
1. Figure 3 shows the resistivity measurement on these samples. As
sample become cleaner (increasing RRR from left to right), there is
a corresponding change in the temperature range (shaded region, from
$T_{\mathrm{low}}$ to $T_{\mathrm{high}}$) over which a linear $\rho(T)$
appears. In cleaner sample, $T_{\mathrm{low}}$ decreases, while the
temperature ratio $T_{\mathrm{high}}/T_{\mathrm{low}}$ over which
we observe linear $\rho(T)$ increases. This behavior is consistent
with Rosch's model \citep{Rosch2000} where $T_{\mathrm{low}}\sim1/k_{F}l$
and $T_{\mathrm{high}}/T_{\mathrm{low}}\sim\sqrt{k_{F}l}$ (see Supplementary
Note 2).}

We found that at lower temperatures ($T\leq T_{\mathrm{low}}$), the
resistivity shows anomalous sublinear temperature dependence. This
is identified as a downward curvature in the resistivity curve, which
is clearly seen in Fig. 3a at $T<10$ K. Similar behaviors in the
resistivity have been observed in heavy fermion Kondo lattice systems.
In CeCo$_{x}$Rh$_{1-x}$In$_{5}$ compounds, the appearance of a
sublinear $\rho(T)$ was associated with the formation of short-range
AFM order together with Kondo coherence, which are intimately related
to an AFM QCP. Short-range AFM order may also emerge in LNO at low
temperatures. This is corroborated by measurements in a magnetic field,
which suppresses weak AFM order. With an applied magnetic field of
9 T, the sublinear part of $\rho(T)$ at low temperatures is mostly
quenched (see Supplementary Figure 8a). As the sample's RRR becomes
higher, both the magnitude and the onset temperature of the sublinear
part of $\rho(T)$ becomes smaller. In our cleanest sample (shown
in Fig. 3e or Fig. 1b), a small sublinear $\rho(T)$ can be identified
only at very low temperatures (< 300 mK), suggesting that pure LNO
is in close proximity to an AFM QCP. We note that a small sublinear
$\rho(T)$ has also been observed at low temperatures in high-quality
bulk single-crystal LNO (Supplementary Figure 8b).

\begin{figure}[!tp]
\includegraphics[width=8cm]{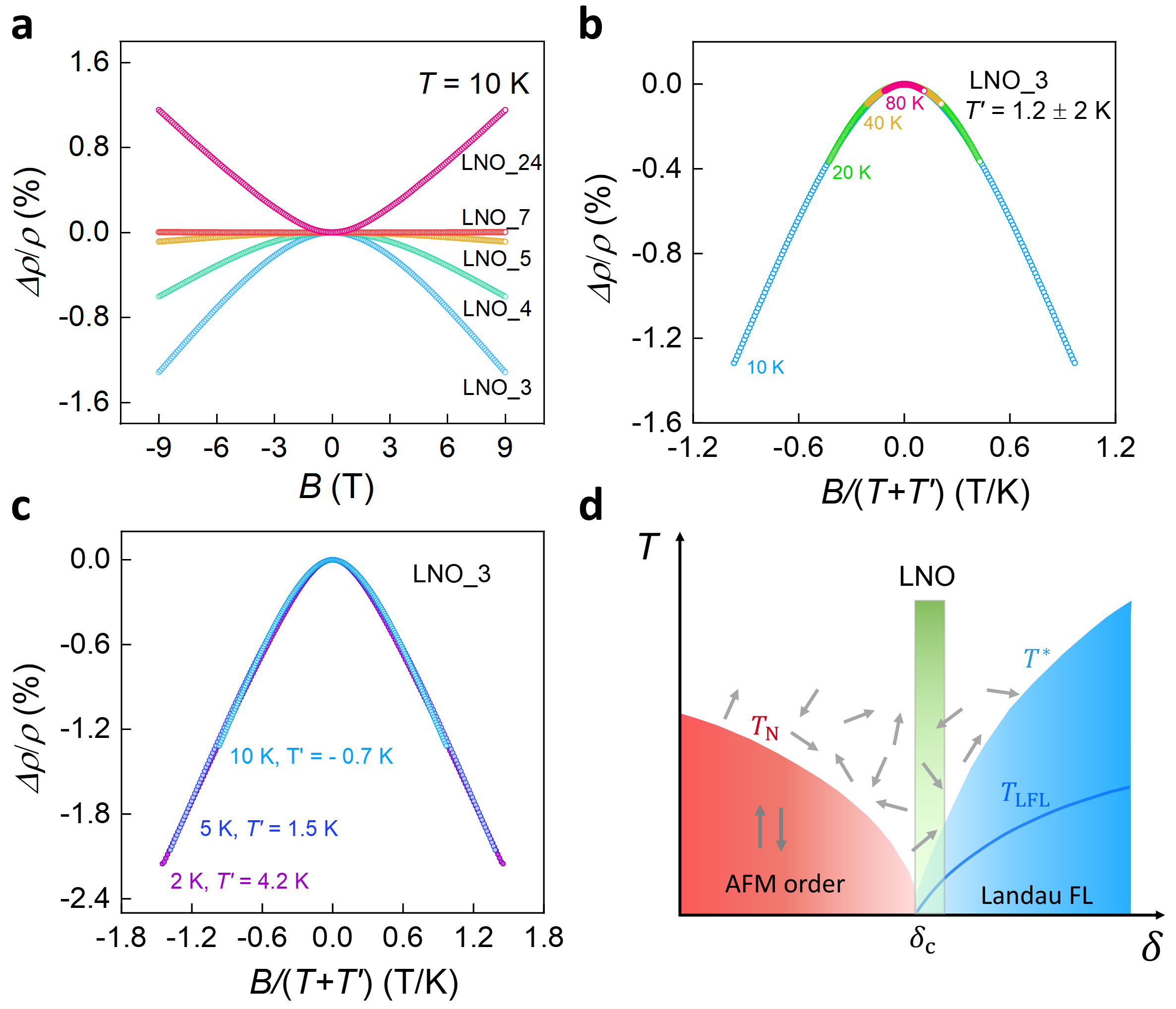}\caption{\textbf{Magnetotransport measurements and phase diagra}\textbf{\textcolor{black}{m.}}\textcolor{black}{{}
}\textbf{\textcolor{black}{a}}\textcolor{black}{{} Longitudinal magnetoresistance
measurements for a series of samples at 10 K. }\textbf{\textcolor{black}{b,
c}}\textcolor{black}{{} The negative LMR measured on LNO\_3 at temperatures
above and below 10 K, respectively. The $x$-axis is scaled by $B/(T+T')$
as discussed in the text. }\textbf{\textcolor{black}{d}}\textcolor{black}{{}
Phase diagram of a system with Kondo-lattice character. Green rectangle
represents the position where LNO is likely to sit. $T^{*}$ marks
the temperature at which Kondo coherence starts to develop. $T_{\mathrm{LFL}}$
represents the temperature below which LFL behavior appears, such
as a quadratic temperature dependence of the resistivity. Small arrows
with random orientations represent local magnetic moments.}}
\end{figure}

\noindent \textbf{Scattering from local magnetic moments}.\textbf{
}A\textcolor{black}{ccording to current understanding, nickelates
are negative charge-transfer materials \citep{Johnston2014,Green2016,Catalano2018}
where the nominally $3d^{7}$-states of the Ni$^{3+}$ sites take
one electron from oxygen leading to a $3d^{8}\underline{L}$ electron
configuration, where $\underline{L}$ stands for a ligand hole on
the O $2p$ orbitals. This may be seen (approximately, due to Ni-O
covalence) as a localized $3d^{8}$ magnetic moment ($S$ = 1) at
a Ni impurity site \citep{Park2012} that is partially screened by
the ligand hole which takes on the $e_{g}$ symmetry of the $3d^{8}$
orbitals. The Fermi level lies in the strongly hybridized O 2p band,
with the lowest lying excitations being from $2p$ (filled) to $2p$
(empty) states in the continuum. Such low lying excitations have been
inferred \citep{Bisogni2016} from resonant X-ray inelastic scattering
measurements in NdNiO$_{3}$, and the dominance of holes in magnetotransport
and thermogalvanic properties has also been experimentally demonst}rated
\citep{Liu2019} recently for LNO.\textbf{\textcolor{red}{{} }}\textcolor{black}{The
screening and hybridization between the conduction charge carriers
and localized magnetic moments in LNO bear resemblance to an underscreened
Kondo lattice} \citep{Lee2011a}. Thus, when an electron is doped
into LNO, it would fill a ligand hole $\underline{L}$ and create
an extra unscreened $3d^{8}$ moment localized on the Ni site, which
can act like a magnetic scattering center.

In our samples, we observe a systematic evolution of the magnetoresistance
with impurity levels. Figure 4a shows our longitudinal magnetoresistance
(LMR) measurement results. The magnetic field is in the film plane
and parallel to the current direction, which minimizes orbital MR
effects from the Lorentz force. The LMR changes gradually from negative
to positive as the RRR increases from 3.3 (LNO\_3) to 24 (LNO\_24).\textcolor{black}{{}
In samples, such as LNO\_3, showing a large negative LMR, the magnitude
of the LMR does not depend on the direction of the magnetic field
(see Supplementary Figure 9). The resistivity also shows metallic
behavior down to the lowest temperature. These phenomena could not
be explained by a weak localization mechanism \citep{Scherwitzl2011}.}\textcolor{blue}{{}
}Furthermore, for $T\geq10$ K, the normalized MR defined as $\Delta\rho/\rho=[(\rho(B,T)-\rho(0,T)]/\rho(0,T)$,
can be made to collapse by plotting $\Delta\rho/\rho$ vs. $B/(T+T')$,
allowing for a variance ($\pm$ 2 K) in the free parameter $T'$,
as shown in Fig. 4(b) (note that $T^{'}\ll T$ in this temperature
range). In particular, $\Delta\rho/\rho$ \textasciitilde{} $-[B/(T+T')]^{2}$,
in agreement with predictions for negative MR induced by spin-flip
scattering from localized impurities \citep{Peski-Tinbergen1963}.
Here, the overall $B/(T+T')$ scaling can be understood as the temperature
dependence of the magnetic susceptibility $\chi$ of single ions;
$T'$ is a measure of correlation among them, analogous to the Curie-\textcolor{black}{Weiss
temperature. At lower temperatures ($T$ < 7 K), shown in Fig. 4(c),
while $\Delta\rho/\rho$ \textasciitilde{} - $B^{2}$ still holds,
the negative LMR begins to saturate, and $T'$ increases as $T$ decreases
approximately as $T'\sim$ (6.2 K - $T$). Essentially, the denominator
in }$B/(T+T')$ remains constant at low temperatures. \textcolor{black}{This
behavior is similar to the Curie-Weiss law for an antiferromagnet,
where the magnetic susceptibility remains finite when the AFM correlation
increases at low temperatures (see Supplementary Figure 10 and 11).
Crucially, the temperature at which the negative LMR starts to saturate
matches the temperature where the $\rho(T)$ becomes sublinear. These
magnetotransport measurements, therefore, further support  that pure
LNO is in the vicinity of an AFM QCP; while the introduction of magnetic
impurities results in short-range AFM ordering (see Supplementary
Note 3). It has been shown that only a small amount of oxygen vacancies
in LNO, which promotes the formation of Ni$^{2+}$ sites, can result
in long-range magnetic ordering \citep{Wang2018}.}

\noindent \textbf{Discussion}

\noindent The transport property of LNO shows similarities to those
reported for Kondo lattice or heavy-fermion materials\citep{Malinowski2005}.
For instance, CeCoIn$_{5}$ shows a linear $\rho(T)$ at low temperatures,
which becomes quadratic in a magnetic field. The magnetic field dependence
of resistivity also shows a $\Delta\rho/\rho$ \textasciitilde{} $-[B/(T+T')]^{2}$
dependence. Thus, the low temperature physics of LNO might involve
a subtle interplay of AFM quantum flu\textcolor{black}{ctuations and
Kondo physics \citep{Doniach1977,Coqblin2006,Si2010,Karner2019a}.
Figure 4d shows} a proposed phase diagram containing LNO. As the tuning
parameter $\delta$ increases, the system undergoes a transition from
an AFM phase (under $T_{\mathrm{N}}$) to a LFL sta\textcolor{black}{te.
Here, $\delta$ can be the $Re$ cation radius or strain. The green
rectangle represents where LNO is likely situated in this phase diagram,
near the }critical point and on its right side. It is understood from
this phase diagram that LNO can be driven from strange metal towards
a LFL by an external magnetic field upon suppressing AFM fluctuations
n\textcolor{black}{ear the QCP. At elevated temperatures, local magnetic
moments give rise to single-impurity spin-flip scattering, particularly
for samples with a higher degree of disorder. Recently, superconductivity
has been observed in an infinite-layer nickelate \citep{Li2019},
which might have a connection to the QCP that we observe here in LNO.
This possibility warrants further exploration.}

\textcolor{black}{In summary, we have observed non-Landau Fermi liquid
behavior at low temperatures in high-purity epitaxial thin films of
LNO. In particular, a linear-in-temperature dependence of resistivity
is observed that extends over almost a decade of temperature at $T\leq1.1$
K in our cleanest samples. The evolution of the resistivity exponent
as a function of temperature follows model predictions f}or a system
with three dimensional quantum critical AFM fluctuations and weak
impurity scattering. These fluctuations are suppressed in a magnetic
field, and the LFL is restored. These results, and the systematics
of single-impurity scattering in samples with elevated disorder level
suggest that high-\textcolor{black}{purity} LNO is on the verge of
an AFM quantum phase transition.

\noindent \textbf{Methods}

\noindent \textbf{Sample growth details}\textcolor{black}{. }When
the effusion cells containing La and Ni sources are heated to deposition
temperatures (\textasciitilde 1430 \textcelsius{} for La and \textasciitilde 1290
\textcelsius{} for Ni), atoms are evaporated towards the substrate
at a stable rate. The deposition rate is measured by a quartz crystal
microbalance (QCM) before and after growth. To calibrate the QCM measurement,
we deposit La and Ni on a MgO substrate using the same growth condition
as for the actual samples. We then use Rutherford backscattering spectrometry
(RBS) to determine the relative ratio of La to Ni. Using a MgO substrate
makes the background near La and Ni peaks clean in the RBS measurement.
Shown in Supplementary Figure 1 is the analysis of the RBS data, which
give a ratio of La to Ni of about 1.009. This value was then used
to adjust the shutter times of La and Ni sources during growth to
target a nominal La/Ni ratio to as close to one as possible. In our
lowest resistivity samples, the drift in La and Ni rates during growth
were under 0.3\% h$^{-1}$.

Supplementary Table 2 presents detailed growth parameters of the samples
discussed in the main text. The growth temperature was measured from
a thermocouple. Elemental sources of La and Ni were evaporated sequentially
from effusion cells using a block-by-block technique. The layer sequence
was LaO - NiO$_{2}$ - LaO.... The growth process was monitored by
reflection high-energy electron diffraction (RHEED). Shown in Supplementary
Figure 2a is the RHEED intensity measured as a function of time as
growth proceeds. When the LaO layer is deposited the surface becomes
rough, which results in a drop in RHEED intensity as shown in Supplementary
Figure 2b. The deposition of a NiO$_{2}$ layer makes the sample surface
smooth enhancing the RHEED intensity, which is shown in Supplementary
Figure 2c. We note that RHEED intensity oscillations can have complex
origins\citep{Sun2018}. Therefore, one period of a RHEED oscillation
corresponds to one unit-cell of LaNiO$_{3}$. In Supplementary Figure
2c, half ordering peaks can be clearly identified, which indicates
good surface crystallinity of the LaNiO$_{3}$ film. We observed that
although the high-RRR samples have good surface crystallinity, the
surface roughness is higher than those low-RRR samples grown at lower
temperatures. Therefore, the sample\textquoteright s conductivity
does not positively correlate with the sample\textquoteright s roughness.
Supplementary Figure 2d shows the characterization of the surface
roughness on sample LNO\_18 using X-ray reflectivity measurement.
The roughness of this high-RRR sample is about 0.86 nm, which is higher
than that of LNO\_3 of about 0.38 nm determined from the same measurement.
After growth, the film's thickness and c-axis parameter were characterized
by low-angle X-ray reflectivity (XRR) and X-ray diffraction (XRD)
measurements, respectively (see Supplementary Figure 3). The film
thickness obtained from fitting the XRR data was \textasciitilde{}
30.8 nm, which is in good agreement (< 1\% error) for a 80-unit cell
LNO with the c-axis lattice constant of \textasciitilde{} 3.82 $\mathrm{\mathring{A}}$
obtained from the XRD. 

\noindent \textbf{Device fabrication and measurement details}\textcolor{black}{.
}Standard photolithography was used to fabricate Hall bar devices
used in the transport measurement. To prevent the formation of oxygen
vacancies, liquid-nitrogen cooling was used during Ar-ion milling.
Electrical contacts were made by depositing 50-nm thick platinum using
sputtering. The channel of the Hall bar has a dimension of 50 x 800
$\mathrm{\mu m^{2}}$, and the distance between the two voltage contacts
is 400 $\mathrm{\mu m}$. Low-noise transport measurements were performed
using a Lakeshore model 372 AC resistance bridge, Stanford lock-in
amplifiers and Keithley 6221/2182A current source/nanovoltmeters in
delta mode. Typical excitation currents of \textasciitilde{} 2 $\mathrm{\mu A}$
were used in our measurements. Cooling of electrons to temperatures
b\textcolor{black}{elow 100 mK was achieved by multi-stage filtering
of the measurement lines in a dilution fridge. }

\noindent \textbf{Uncertainties of the measurement}\textcolor{black}{.
In all figures including those in the Supplementary Information, the
data points for the resistivity measurement are interpolations or
mean values of the raw data which have higher density than those shown
in each plot. The error bar, representing the standard deviation of
the mean at each data point, is smaller than the symbol size. In these
measurements, the uncertainty in resistivity is approximately $8\times10^{-4}$
$\mathrm{\mu}\Omega$ cm. There is a systematic uncertainty (< 10\%)
in resistivity that comes from variations in the dimensions of individual
device, which does not affect the RRR value of the sample.}

\noindent \textbf{Data Availability} 

\noindent The data that support the findings of this study are available
from the corresponding author upon reasonable request.

\noindent \textbf{Acknowledgment}

\noindent All work at Argonne including film growth, characterization
and magnetotransport measurements were supported by the US Department
of Energy, Office of Science, Basic Energy Sciences, Materials Sciences
and Engineering Division including support for F.W. via the Center
for Predictive Simulation of Functional Materials. The use of facilities
at the Center for Nanoscale Materials, an Office of Science user facility,
was supported by the US Department of Energy, Basic Energy Sciences
under Contract No. DE-AC02-06CH11357. Dilution fridge measurements
carried out by G. W., J. Z and C. C. at the Argonne High Energy Physics
Division was supported by the US Department of Energy, Office of Science,
High Energy Physics. Dilution fridge measurements at UIUC were carried
out by V. F. C. H. and N. M., supported by the National Science Foundation
under NSF DMR- 1710437. Measurements at low temperatures and high
magnetic fields were carried out at the NHMFL, which is supported
by the NSF Cooperative agreement no. DMR-1157490 and the State of
Florida. We thank Peter Littlewood, John Mitchell and Andrew Millis
for discussions.

\noindent \textbf{Author contributions }

\noindent C. L. grew and characterized the LNO samples with assistance
from D. H., J. D. H. and F. W.. Transport measurements were carried
out by C. L., V. F. C. H., T. B. S., G. W., J. Z., J. S. J., C. C.,
A. S. and N. M., down to dilution fridge temperatures. M. R. N. provided
theoretical guidance. C. L. and A. B. wrote the manuscript with contributions
from all authors. A. B. supervised the project.

\noindent \textbf{Competing interests}

\noindent The authors declare no competing financial interests.

\begin{thebibliography}{10}
\expandafter\ifx\csname url\endcsname\relax
  \def\url#1{\texttt{#1}}\fi
\expandafter\ifx\csname urlprefix\endcsname\relax\def\urlprefix{URL }\fi
\providecommand{\bibinfo}[2]{#2}
\providecommand{\eprint}[2][]{\url{#2}}

\bibitem{Coleman2005}
\bibinfo{author}{Coleman, P.} \& \bibinfo{author}{Schofield, A.~J.}
\newblock \bibinfo{title}{Quantum criticality}.
\newblock \emph{\bibinfo{journal}{Nature}} \textbf{\bibinfo{volume}{433}},
  \bibinfo{pages}{226--229} (\bibinfo{year}{2005}).

\bibitem{Gegenwart2008}
\bibinfo{author}{Gegenwart, P.}, \bibinfo{author}{Si, Q.} \&
  \bibinfo{author}{Steglich, F.}
\newblock \bibinfo{title}{Quantum criticality in heavy-fermion metals}.
\newblock \emph{\bibinfo{journal}{Nat. Phys.}} \textbf{\bibinfo{volume}{4}},
  \bibinfo{pages}{186--197} (\bibinfo{year}{2008}).

\bibitem{Sachdev2009}
\bibinfo{author}{Sachdev, S.}
\newblock \emph{\bibinfo{title}{Quantum Phase Transitions}}
  (\bibinfo{publisher}{Cambridge University Press}, \bibinfo{year}{2009}).

\bibitem{Schofield1999}
\bibinfo{author}{Schofield, A.~J.}
\newblock \bibinfo{title}{Non-fermi liquids}.
\newblock \emph{\bibinfo{journal}{Contemp. Phys.}}
  \textbf{\bibinfo{volume}{40}}, \bibinfo{pages}{95--115}
  (\bibinfo{year}{1999}).

\bibitem{Stewart2001}
\bibinfo{author}{Stewart, G.~R.}
\newblock \bibinfo{title}{Non-fermi-liquid behavior in $d$- and $f$-electron
  metals}.
\newblock \emph{\bibinfo{journal}{Rev. Mod. Phys.}}
  \textbf{\bibinfo{volume}{73}}, \bibinfo{pages}{797--855}
  (\bibinfo{year}{2001}).

\bibitem{Torrance1992}
\bibinfo{author}{Torrance, J.~B.}, \bibinfo{author}{Lacorre, P.},
  \bibinfo{author}{Nazzal, A.~I.}, \bibinfo{author}{Ansaldo, E.~J.} \&
  \bibinfo{author}{Niedermayer, C.}
\newblock \bibinfo{title}{Systematic study of insulator-metal transitions in
  perovskites {${\textit{R}}{\mathrm{NiO}}_{3}$} {(${\textit{R}}$ = Pr, Nd, Sm,
  Eu)} due to closing of charge-transfer gap}.
\newblock \emph{\bibinfo{journal}{Phys. Rev. B}} \textbf{\bibinfo{volume}{45}},
  \bibinfo{pages}{8209--8212} (\bibinfo{year}{1992}).

\bibitem{Catalano2018}
\bibinfo{author}{Catalano, S.} \emph{et~al.}
\newblock \bibinfo{title}{Rare-earth nickelates
  {${\textit{R}}{\text{NiO}}_{3}$} : thin films and heterostructures}.
\newblock \emph{\bibinfo{journal}{Rep. Prog. Phys.}}
  \textbf{\bibinfo{volume}{81}}, \bibinfo{pages}{046501}
  (\bibinfo{year}{2018}).

\bibitem{Zhou2014}
\bibinfo{author}{Zhou, J.-S.}, \bibinfo{author}{Marshall, L.~G.} \&
  \bibinfo{author}{Goodenough, J.~B.}
\newblock \bibinfo{title}{Mass enhancement versus stoner enhancement in
  strongly correlated metallic perovskites: {LaNiO$_{3}$} and {LaCuO$_{3}$}}.
\newblock \emph{\bibinfo{journal}{Phys. Rev. B}} \textbf{\bibinfo{volume}{89}},
  \bibinfo{pages}{245138} (\bibinfo{year}{2014}).

\bibitem{Hoffman2016}
\bibinfo{author}{Hoffman, J.~D.} \emph{et~al.}
\newblock \bibinfo{title}{Oscillatory noncollinear magnetism induced by
  interfacial charge transfer in superlattices composed of metallic oxides}.
\newblock \emph{\bibinfo{journal}{Phys. Rev. X}} \textbf{\bibinfo{volume}{6}},
  \bibinfo{pages}{041038} (\bibinfo{year}{2016}).

\bibitem{Fabbris2018}
\bibinfo{author}{Fabbris, G.} \emph{et~al.}
\newblock \bibinfo{title}{Emergent $c$-axis magnetic helix in
  manganite-nickelate superlattices}.
\newblock \emph{\bibinfo{journal}{Phys. Rev. B}} \textbf{\bibinfo{volume}{98}},
  \bibinfo{pages}{180401} (\bibinfo{year}{2018}).

\bibitem{Allen2015}
\bibinfo{author}{Allen, S.~J.} \emph{et~al.}
\newblock \bibinfo{title}{Gaps and pseudogaps in perovskite rare earth
  nickelates}.
\newblock \emph{\bibinfo{journal}{APL Mater.}} \textbf{\bibinfo{volume}{3}},
  \bibinfo{pages}{062503} (\bibinfo{year}{2015}).

\bibitem{Subedi2018}
\bibinfo{author}{Subedi, A.}
\newblock \bibinfo{title}{Breathing distortions in the metallic,
  antiferromagnetic phase of {LaNiO}${_{3}}$}.
\newblock \emph{\bibinfo{journal}{SciPost Phys.}} \textbf{\bibinfo{volume}{5}},
  \bibinfo{pages}{20} (\bibinfo{year}{2018}).

\bibitem{Zhang2017}
\bibinfo{author}{Zhang, J.}, \bibinfo{author}{Zheng, H.}, \bibinfo{author}{Ren,
  Y.} \& \bibinfo{author}{Mitchell, J.~F.}
\newblock \bibinfo{title}{High-pressure floating-zone growth of perovskite
  nickelate {LaNiO}${_{3}}$ single crystals}.
\newblock \emph{\bibinfo{journal}{Cryst. Growth Des.}}
  \textbf{\bibinfo{volume}{17}}, \bibinfo{pages}{2730--2735}
  (\bibinfo{year}{2017}).

\bibitem{Rivadulla2003}
\bibinfo{author}{Rivadulla, F.}, \bibinfo{author}{Zhou, J.-S.} \&
  \bibinfo{author}{Goodenough, J.~B.}
\newblock \bibinfo{title}{Electron scattering near an itinerant to localized
  electronic transition}.
\newblock \emph{\bibinfo{journal}{Phys. Rev. B}} \textbf{\bibinfo{volume}{67}},
  \bibinfo{pages}{165110} (\bibinfo{year}{2003}).

\bibitem{Guo2018}
\bibinfo{author}{Guo, H.} \emph{et~al.}
\newblock \bibinfo{title}{Antiferromagnetic correlations in the metallic
  strongly correlated transition metal oxide {LaNiO${}_{3}$}}.
\newblock \emph{\bibinfo{journal}{Nat. Commun.}} \textbf{\bibinfo{volume}{9}},
  \bibinfo{pages}{43} (\bibinfo{year}{2018}).

\bibitem{Liu2013}
\bibinfo{author}{Liu, J.} \emph{et~al.}
\newblock \bibinfo{title}{Heterointerface engineered electronic and magnetic
  phases of {NdNiO$_{3}$} thin films}.
\newblock \emph{\bibinfo{journal}{Nat. Commun.}} \textbf{\bibinfo{volume}{4}}
  (\bibinfo{year}{2013}).

\bibitem{Mikheev2015}
\bibinfo{author}{Mikheev, E.} \emph{et~al.}
\newblock \bibinfo{title}{Tuning bad metal and non-fermi liquid behavior in a
  mott material: Rare-earth nickelate thin films}.
\newblock \emph{\bibinfo{journal}{Sci. Adv.}} \textbf{\bibinfo{volume}{1}},
  \bibinfo{pages}{e1500797} (\bibinfo{year}{2015}).

\bibitem{moriya2012spin}
\bibinfo{author}{Moriya, T.}
\newblock \emph{\bibinfo{title}{Spin fluctuations in itinerant electron
  magnetism}}, vol.~\bibinfo{volume}{56} (\bibinfo{publisher}{Springer Science
  \& Business Media}, \bibinfo{year}{2012}).

\bibitem{Millis1993}
\bibinfo{author}{Millis, A.~J.}
\newblock \bibinfo{title}{Effect of a nonzero temperature on quantum critical
  points in itinerant fermion systems}.
\newblock \emph{\bibinfo{journal}{Phys. Rev. B}} \textbf{\bibinfo{volume}{48}},
  \bibinfo{pages}{7183--7196} (\bibinfo{year}{1993}).

\bibitem{LohneyesenRMP2007}
\bibinfo{author}{L\"ohneysen, H.~v.}, \bibinfo{author}{Rosch, A.},
  \bibinfo{author}{Vojta, M.} \& \bibinfo{author}{W\"olfle, P.}
\newblock \bibinfo{title}{Fermi-liquid instabilities at magnetic quantum phase
  transitions}.
\newblock \emph{\bibinfo{journal}{Rev. Mod. Phys.}}
  \textbf{\bibinfo{volume}{79}}, \bibinfo{pages}{1015--1075}
  (\bibinfo{year}{2007}).

\bibitem{Hlubina1995}
\bibinfo{author}{Hlubina, R.} \& \bibinfo{author}{Rice, T.~M.}
\newblock \bibinfo{title}{Resistivity as a function of temperature for models
  with hot spots on the fermi surface}.
\newblock \emph{\bibinfo{journal}{Phys. Rev. B}} \textbf{\bibinfo{volume}{51}},
  \bibinfo{pages}{9253--9260} (\bibinfo{year}{1995}).

\bibitem{Rosch1999}
\bibinfo{author}{Rosch, A.}
\newblock \bibinfo{title}{Interplay of disorder and spin fluctuations in the
  resistivity near a quantum critical point}.
\newblock \emph{\bibinfo{journal}{Phys. Rev. Lett.}}
  \textbf{\bibinfo{volume}{82}}, \bibinfo{pages}{4280--4283}
  (\bibinfo{year}{1999}).

\bibitem{Rosch2000}
\bibinfo{author}{Rosch, A.}
\newblock \bibinfo{title}{Magnetotransport in nearly antiferromagnetic metals}.
\newblock \emph{\bibinfo{journal}{Phys. Rev. B}} \textbf{\bibinfo{volume}{62}},
  \bibinfo{pages}{4945--4962} (\bibinfo{year}{2000}).

\bibitem{Custers2003}
\bibinfo{author}{Custers, J.} \emph{et~al.}
\newblock \bibinfo{title}{The break-up of heavy electrons at a quantum critical
  point}.
\newblock \emph{\bibinfo{journal}{Nature}} \textbf{\bibinfo{volume}{424}},
  \bibinfo{pages}{524} (\bibinfo{year}{2003}).

\bibitem{Paglione2003}
\bibinfo{author}{Paglione, J.} \emph{et~al.}
\newblock \bibinfo{title}{Field-induced quantum critical point in
  {${\mathrm{C}\mathrm{e}\mathrm{C}\mathrm{o}\mathrm{I}\mathrm{n}}_{5}$}}.
\newblock \emph{\bibinfo{journal}{Phys. Rev. Lett.}}
  \textbf{\bibinfo{volume}{91}}, \bibinfo{pages}{246405}
  (\bibinfo{year}{2003}).

\bibitem{Son2010}
\bibinfo{author}{Son, J.} \emph{et~al.}
\newblock \bibinfo{title}{Low-dimensional mott material: Transport in ultrathin
  epitaxial {LaNiO$_{3}$} films}.
\newblock \emph{\bibinfo{journal}{Appl. Phys. Lett.}}
  \textbf{\bibinfo{volume}{96}}, \bibinfo{pages}{062114}
  (\bibinfo{year}{2010}).

\bibitem{Johnston2014}
\bibinfo{author}{Johnston, S.}, \bibinfo{author}{Mukherjee, A.},
  \bibinfo{author}{Elfimov, I.}, \bibinfo{author}{Berciu, M.} \&
  \bibinfo{author}{Sawatzky, G.~A.}
\newblock \bibinfo{title}{Charge disproportionation without charge transfer in
  the rare-earth-element nickelates as a possible mechanism for the
  metal-insulator transition}.
\newblock \emph{\bibinfo{journal}{Phys. Rev. Lett.}}
  \textbf{\bibinfo{volume}{112}}, \bibinfo{pages}{106404}
  (\bibinfo{year}{2014}).

\bibitem{Green2016}
\bibinfo{author}{Green, R.~J.}, \bibinfo{author}{Haverkort, M.~W.} \&
  \bibinfo{author}{Sawatzky, G.~A.}
\newblock \bibinfo{title}{Bond disproportionation and dynamical charge
  fluctuations in the perovskite rare-earth nickelates}.
\newblock \emph{\bibinfo{journal}{Phys. Rev. B}} \textbf{\bibinfo{volume}{94}},
  \bibinfo{pages}{195127} (\bibinfo{year}{2016}).

\bibitem{Park2012}
\bibinfo{author}{Park, H.}, \bibinfo{author}{Millis, A.~J.} \&
  \bibinfo{author}{Marianetti, C.~A.}
\newblock \bibinfo{title}{Site-selective mott transition in rare-earth-element
  nickelates}.
\newblock \emph{\bibinfo{journal}{Phys. Rev. Lett.}}
  \textbf{\bibinfo{volume}{109}} (\bibinfo{year}{2012}).

\bibitem{Bisogni2016}
\bibinfo{author}{Bisogni, V.} \emph{et~al.}
\newblock \bibinfo{title}{Ground-state oxygen holes and the metal-insulator
  transition in the negative charge-transfer rare-earth nickelates}.
\newblock \emph{\bibinfo{journal}{Nat. Commun.}} \textbf{\bibinfo{volume}{7}},
  \bibinfo{pages}{13017} (\bibinfo{year}{2016}).

\bibitem{Liu2019}
\bibinfo{author}{Liu, C.} \emph{et~al.}
\newblock \bibinfo{title}{Counter-thermal flow of holes in high-mobility
  {${\mathrm{LaNiO}}_{3}$} thin films}.
\newblock \emph{\bibinfo{journal}{Phys. Rev. B}} \textbf{\bibinfo{volume}{99}},
  \bibinfo{pages}{041114} (\bibinfo{year}{2019}).

\bibitem{Lee2011a}
\bibinfo{author}{Lee, S.}, \bibinfo{author}{Chen, R.} \&
  \bibinfo{author}{Balents, L.}
\newblock \bibinfo{title}{Metal-insulator transition in a two-band model for
  the perovskite nickelates}.
\newblock \emph{\bibinfo{journal}{Phys. Rev. B}} \textbf{\bibinfo{volume}{84}},
  \bibinfo{pages}{165119} (\bibinfo{year}{2011}).

\bibitem{Scherwitzl2011}
\bibinfo{author}{Scherwitzl, R.} \emph{et~al.}
\newblock \bibinfo{title}{Metal-insulator transition in ultrathin {LaNiO$_{3}$}
  films}.
\newblock \emph{\bibinfo{journal}{Phys. Rev. Lett.}}
  \textbf{\bibinfo{volume}{106}} (\bibinfo{year}{2011}).

\bibitem{Peski-Tinbergen1963}
\bibinfo{author}{Peski-Tinbergen, T.~V.} \& \bibinfo{author}{Dekker, A.}
\newblock \bibinfo{title}{Spin-dependent scattering and resistivity of magnetic
  metals and alloys}.
\newblock \emph{\bibinfo{journal}{Physica}} \textbf{\bibinfo{volume}{29}},
  \bibinfo{pages}{917--937} (\bibinfo{year}{1963}).

\bibitem{Wang2018}
\bibinfo{author}{Wang, B.-X.} \emph{et~al.}
\newblock \bibinfo{title}{Antiferromagnetic defect structure in
  {$\mathrm{LaNi}{\mathrm{O}}_{3\ensuremath{-}\ensuremath{\delta}}$} single
  crystals}.
\newblock \emph{\bibinfo{journal}{Phys. Rev. Materials}}
  \textbf{\bibinfo{volume}{2}}, \bibinfo{pages}{064404} (\bibinfo{year}{2018}).

\bibitem{Malinowski2005}
\bibinfo{author}{Malinowski, A.} \emph{et~al.}
\newblock \bibinfo{title}{c-axis magnetotransport in {${\text{CeCoIn}}_{5}$}}.
\newblock \emph{\bibinfo{journal}{Phys. Rev. B}} \textbf{\bibinfo{volume}{72}}
  (\bibinfo{year}{2005}).

\bibitem{Doniach1977}
\bibinfo{author}{Doniach, S.}
\newblock \bibinfo{title}{The kondo lattice and weak antiferromagnetism}.
\newblock \emph{\bibinfo{journal}{Physica B+C}} \textbf{\bibinfo{volume}{91}},
  \bibinfo{pages}{231 -- 234} (\bibinfo{year}{1977}).

\bibitem{Coqblin2006}
\bibinfo{author}{Coqblin, B.}, \bibinfo{author}{ez~Regueiro, M. D.~N.},
  \bibinfo{author}{Theumann, A.}, \bibinfo{author}{Iglesias, J.~R.} \&
  \bibinfo{author}{Magalhães, S.~G.}
\newblock \bibinfo{title}{Theory of the kondo lattice: competition between
  kondo effect and magnetic order}.
\newblock \emph{\bibinfo{journal}{Philos. Mag.}} \textbf{\bibinfo{volume}{86}},
  \bibinfo{pages}{2567--2580} (\bibinfo{year}{2006}).

\bibitem{Si2010}
\bibinfo{author}{Si, Q.} \& \bibinfo{author}{Steglich, F.}
\newblock \bibinfo{title}{Heavy fermions and quantum phase transitions}.
\newblock \emph{\bibinfo{journal}{Science}} \textbf{\bibinfo{volume}{329}},
  \bibinfo{pages}{1161--1166} (\bibinfo{year}{2010}).

\bibitem{Karner2019a}
\bibinfo{author}{Karner, V.~L.} \emph{et~al.}
\newblock \bibinfo{title}{Local metallic and structural properties of the
  strongly correlated metal $\mathrm{LaNiO}{}_{3}$ using ${}^{8}\mathrm{Li}$
  $\ensuremath{\beta}\text{--}\mathrm{NMR}$}.
\newblock \emph{\bibinfo{journal}{Phys. Rev. B}}
  \textbf{\bibinfo{volume}{100}}, \bibinfo{pages}{165109}
  (\bibinfo{year}{2019}).

\bibitem{Li2019}
\bibinfo{author}{Li, D.} \emph{et~al.}
\newblock \bibinfo{title}{Superconductivity in an infinite-layer nickelate}.
\newblock \emph{\bibinfo{journal}{Nature}} \textbf{\bibinfo{volume}{572}},
  \bibinfo{pages}{624--627} (\bibinfo{year}{2019}).

\bibitem{Sun2018}
\bibinfo{author}{Sun, H.~Y.} \emph{et~al.}
\newblock \bibinfo{title}{Chemically specific termination control of oxide
  interfaces via layer-by-layer mean inner potential engineering}.
\newblock \emph{\bibinfo{journal}{Nat. Commun.}} \textbf{\bibinfo{volume}{9}},
  \bibinfo{pages}{2965} (\bibinfo{year}{2018}).

\end{thebibliography}
\end{document}